\documentclass[11pt]{article}
\usepackage{amssymb,amsmath,amsfonts}
\usepackage{graphicx}
\usepackage{graphics}
\usepackage{eepic,epsfig}
\usepackage{dcolumn}

1.60cm \makeatletter \@addtoreset{equation}{section}

\makeatother

\begin{document}

\title{Wightman function and Casimir densities for Robin plates \\
in the Fulling--Rindler vacuum }
\author{A. A. Saharian\thanks{%
Email: saharyan@server.physdep.r.am}, R. M. Avagyan, and R. S. Davtyan \\
\textit{Department of Physics, Yerevan State University, 1 Alex Manoogian
Str.} \\
\textit{\ 375049 Yerevan, Armenia}}
\date{\today }
\maketitle

\begin{abstract}
Wightman function, the vacuum expectation values of the field square
and the energy--momentum tensor are investigated for a massive
scalar field with an arbitrary curvature coupling parameter in the
region between two infinite parallel plates moving by uniform proper
acceleration. We assume that the field is prepared in the
Fulling--Rindler vacuum state and satisfies Robin boundary
conditions on the plates. The mode--summation method is used with a
combination of a variant of the generalized Abel--Plana formula.
This allows to extract manifestly the contributions to the
expectation values due to a single boundary and to present the
second plate-induced parts in terms of exponentially convergent
integrals. Various limiting cases are investigated. The vacuum
forces acting on the boundaries are presented as a sum of the
self--action and 'interaction' terms. The first one contains well
known surface divergences and needs a further renormalization. The
'interaction' forces between the plates are investigated as
functions of the proper accelerations and coefficients in the
boundary conditions. We show that there is a region in the space of
these parameters in which the 'interaction' forces are repulsive for
small distances and attractive for large distances.
\end{abstract}

\bigskip

PACS number(s): 03.70.+k, 11.10.Kk, 04.62.+v

\bigskip

\section{Introduction}

The Casimir effect is one of the most interesting macroscopic manifestations
of the nontrivial structure of the vacuum state in quantum field theory
(see, e.g., \cite{Mostepanenko,Plunien,Milton,Bordag1} and references
therein). The effect is a phenomenon common to all systems characterized by
fluctuating quantities and results from changes in the vacuum fluctuations
of a quantum field that occur because of the imposition of boundary
conditions or the choice of topology. It may have important implications on
all scales, from cosmological to subnuclear, and has become in recent
decades an increasingly popular topic in quantum field theory. It is well
known that the uniqueness of the vacuum state is lost when we work within
the framework of quantum field theory in a general curved spacetime or in
non--inertial frames. In particular, the use of general coordinate
transformations in quantum field theory in flat spacetime leads to an
infinite number of unitary inequivalent representations of the commutation
relations. Different inequivalent representations will in general give rise
to different vacuum states. For instance, the vacuum state for a uniformly
accelerated observer, the Fulling--Rindler vacuum \cite%
{Full73,Boul75,Unru76,Full77,Gerl89}, turns out to be inequivalent to that
for an inertial observer, the familiar Minkowski vacuum. Quantum field
theory in accelerated systems contains many special features produced by a
gravitational field. In particular, the near horizon geometry of most black
holes is well approximated by Rindler spacetime and a better understanding
of physical effects in this background could serve as a handle to deal with
more complicated geometries like Schwarzschild. The Rindler geometry shares
most of the qualitative features of black holes and is simple enough to
allow detailed analysis. Another motivation for the investigation of quantum
effects in the Rindler space is related to the fact that this space is
conformally related to de Sitter space and to Robertson--Walker space with
negative spatial curvature. As a result the expectation values of the
energy--momentum tensor for conformally invariant fields and for
corresponding conformally transformed boundaries on the de Sitter and
Robertson--Walker backgrounds can be generated from the corresponding
Rindler counterpart by the standard transformation (see, for instance, \cite%
{Birrell}).

An interesting topic in the investigations of the Casimir effect is
the dependence of the vacuum characteristics on the type of the
vacuum. Vacuum expectation values of the energy-momentum tensor
induced by an infinite plane boundary moving with uniform proper
acceleration through the Fulling-Rindler vacuum was studied by
Candelas and Deutsch \cite{Candelas} for the conformally coupled
$4D$ Dirichlet and Neumann massless scalar and electromagnetic
fields. In this paper only the region of the right Rindler wedge to
the right of the barrier is considered. In Ref. \cite{Saha02} we
have investigated the Wightman function and the vacuum
energy-momentum tensor for a massive scalar field with general
curvature coupling parameter, satisfying the Robin boundary
conditions on the infinite plane in an arbitrary number of spacetime
dimensions and for the electromagnetic field. We have considered
both regions, including the one between the barrier and Rindler
horizon. The vacuum expectation values of the energy-momentum tensor
for scalar fields with Dirichlet and Neumann boundary conditions and
for the electromagnetic field in the geometry of two parallel plates
moving by uniform accelerations are investigated in Ref.
\cite{Avag02}. In particular, the vacuum forces acting on the
boundaries are evaluated. They are presented as a sum of the
'interaction' and self-action parts. The 'interaction' forces
between the plates are always attractive for both scalar and
electromagnetic cases. Due to the well-known surface divergences in
the boundary parts, the total Casimir energy cannot be obtained by
direct integration of the vacuum energy density and needs an
additional renormalization. In Ref. \cite{Saha04} by using the zeta
function technique, the Casimir energy is evaluated for massless
scalar fields under Dirichlet and Neumann boundary conditions, and
for the electromagnetic field with perfect conductor boundary
conditions on one and two parallel plates. On background of
manifolds with boundaries, the physical quantities, in general, will
receive both volume and surface contributions and the surface terms
play an important role in various branches of physics. An expression
for the surface energy-momentum tensor for a scalar field with
general curvature coupling parameter in the general case of bulk and
boundary geometries is derived in Ref. \cite{Saha04c}. In Ref.
\cite{SahSet04b} the vacuum expectation value of the surface
energy-momentum tensor is evaluated for a massles scalar field
obeying Robin boundary condition on an infinite plane moving by
uniform proper acceleration. By using the conformal relation between
the Rindler and de Sitter spacetimes and the results from
\cite{Saha02}, in Ref. \cite{SahSet04} the vacuum energy-momentum
tensor for a scalar field is evaluated in de Sitter spacetime in
presence of a curved brane on which the field obeys the Robin
boundary condition with coordinate dependent coefficients.

In the present paper the Wightman function and the vacuum
expectation value of the field square the energy-momentum tensor are
investigated for a massive scalar field with an arbitrary curvature
coupling parameter obeying the Robin boundary conditions on two
parallel branes moving by uniform proper accelerations through the
Fulling-Rindler vacuum. The general case is considered when the
constants in the boundary conditions are different for separate
plates. Robin type conditions are an extension of Dirichlet and
Neumann boundary conditions and appear in a variety of situations,
including the considerations of vacuum effects for a confined
charged scalar field in external fields \cite{Ambj83}, spinor and
gauge field theories, quantum gravity and supergravity
\cite{Luck91,Espo97}. Robin conditions can be made conformally
invariant, while purely-Neumann conditions cannot. Thus, Robin-type
conditions are needed when one deals with conformally invariant
theories in the presence of boundaries and wishes to preserve this
invariance. It is interesting to note that the quantum scalar field
satisfying the Robin condition on the boundary of cavity violates
the Bekenstein's entropy-to-energy bound near certain points in the
space of the parameter defining the boundary condition
\cite{Solo01}. The Robin boundary conditions are an extension of
those imposed on perfectly conducting boundaries and may, in some
geometries, be useful for depicting the finite penetration of the
field into the boundary with the 'skin-depth' parameter related to
the Robin coefficient. Robin boundary conditions naturally arise
for scalar and fermion bulk fields in the Randall-Sundrum model \cite%
{Gher00,Flac01b,Saha05}. In this model the bulk geometry is a slice of
anti-de Sitter space and the corresponding Robin coefficients are related to
the curvature scale of the space.

The outline of this paper is the following. In the next section the
Wightman function is considered. The corresponding mode-sum is
evaluated by using the generalized Abel-Plana summation formula
\cite{Sahrev}. This allows us to extract from the corresponding
vacuum expectation values the Wightman function for the geometry of
a single plate and to present the remained part in the form of the
exponentially convergent integrals. The vacuum expectation values of
the field square and the Casimir energy-momentum tensor is evaluated
in Section \ref{sec:VEVEMT}. Various limiting cases are considered.
In Section \ref{sec:IntForce} we investigate the vacuum
'interaction' forces between the plates as functions on
corresponding proper accelerations. Section \ref{sec:Conc} contains
a summary of the work and some suggestions for further research. In
Appendix \ref{section:App1} on the base of the generalized
Abel-Plana formula, a summation formula is derived for the series
over zeros of a combination of the Bessel modified functions with an
imaginary order.

\section{Wightman function}

\label{sec:WF}

We consider a real scalar field $\varphi (x)$ with general curvature
coupling parameter $\zeta $ satisfying the field equation
\begin{equation}
\left( \nabla _{\mu }\nabla ^{\mu }+m^{2}+\zeta R\right) \varphi =0,
\label{fieldeq}
\end{equation}%
where $R$ is the scalar curvature for a $(D+1)$--dimensional background
spacetime, and $\nabla _{\mu }$ is the covariant derivative operator. For
special cases of minimally and conformally coupled scalars one has $\zeta =0$
and $\zeta =(D-1)/4D$, respectively. Our main interest in this paper will be
the Wightman function, the vacuum expectation values (VEVs) of the field
square and the energy-momentum tensor in the Rindler spacetime induced by
two parallel plates moving with uniform proper acceleration when the quantum
field is prepared in the Fulling-Rindler vacuum. For this problem the
background spacetime is flat and in Eq. (\ref{fieldeq}) we have $R=0$. As a
result the eigenmodes are independent on the curvature coupling parameter.
However, the local characteristics of the vacuum such as energy density and
vacuum stresses \ depend on this parameter.

\begin{figure}[tbph]
\begin{center}
\epsfig{figure=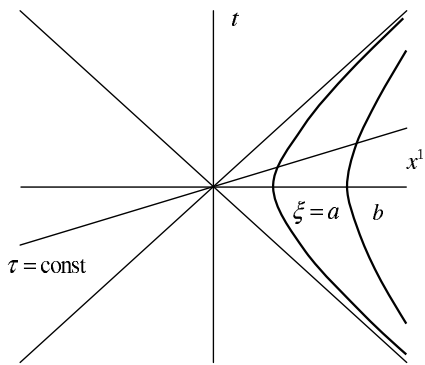,width=6cm,height=5cm}
\end{center}
\caption{The $(x^{1},t)$ plane with the Rindler coordinates. The heavy lines
$\protect\xi =a$ and $\protect\xi = b$ represent the trajectories of the
plates.}
\label{fig1avsa}
\end{figure}

In the accelerated frame it is convenient to introduce Rindler coordinates $%
(\tau ,\xi ,\mathbf{x})$ related to the Minkowski ones, $(t,x^{1},\mathbf{x}%
) $ by formulas
\begin{equation}
t=\xi \sinh \tau ,\quad x^{1}=\xi \cosh \tau ,  \label{RindMin}
\end{equation}%
where $\mathbf{x}=(x^{2},\ldots ,x^{D})$ denotes the set of coordinates
parallel to the plates. In these coordinates the Minkowski line element
takes the form
\begin{equation}
ds^{2}=\xi ^{2}d\tau ^{2}-d\xi ^{2}-d\mathbf{x}^{2},  \label{metric}
\end{equation}%
and a world line defined by $\xi ,\mathbf{x}=\mathrm{const}$ describes an
observer with constant proper acceleration $\xi ^{-1}$. The Rindler time
coordinate $\tau $ is proportional to the proper time along a family of
uniformly accelerated trajectories which fill the Rindler wedge, with the
proportionality constant equal to the acceleration. Assuming that the plates
are situated in the right Rindler wedge $x^{1}>\left\vert t\right\vert $, we
will let the surfaces $\xi =a$ and $\xi =b$, $b>a$ represent the
trajectories of these boundaries, which therefore have proper accelerations $%
a^{-1}$ and $b^{-1}$ (see Fig. \ref{fig1avsa}). We will consider the case of
a scalar field satisfying Robin boundary conditions on the surfaces of the
plates:
\begin{equation}
\left. \left( A_{j}+B_{j}\frac{\partial }{\partial \xi }\right) \varphi
\right\vert _{\xi =j}=0,\quad j=a,b,  \label{Dboundcond}
\end{equation}%
with constant coefficients $A_{j}$ and $B_{j}$. Dirichlet and Neumann
boundary conditions are obtained from here as special cases. All results
below will depend, of course, on the ratios $A_{j}/B_{j}$ only. However, to
keep the transition to Dirichlet and Neumann cases transparent, we write the
boundary conditions in the form (\ref{Dboundcond}).

The plates divide the right Rindler wedge into three regions: $0<\xi <a$, $%
\xi >b$, and $a<\xi <b$. The VEVs in two first regions are the same as those
induced by single plates located at $\xi =a$ and $\xi =b$, respectively. As
these VEVs are investigated in Ref. \cite{Saha02}, in the consideration
below we restrict ourselves to the region between the plates. First we
consider the positive frequency Wightman function $G^{+}(x,x^{\prime
})=\left\langle 0\left\vert \varphi (x)\varphi (x^{\prime })\right\vert
0\right\rangle $, with $\left. |0\right\rangle $ being the amplitude for the
corresponding vacuum state. The VEVs of the field square and the
energy-momentum tensor can be evaluated on the base of this function. In
addition, the Wightman function determines the response of the particle
detector of Unruh-DeWitt type, moving through the vacuum under
consideration. By expanding the field operator over the complete set of
eigenfunctions $\left\{ \varphi _{\alpha }(x),\varphi _{\alpha }^{\ast
}(x)\right\} $ satisfying boundary conditions (\ref{Dboundcond}) and using
the commutation relations one finds
\begin{equation}
G^{+}(x,x^{\prime })=\sum_{\alpha }\varphi _{\alpha }(x)\varphi _{\alpha
}^{\ast }(x^{\prime }),  \label{mswf}
\end{equation}%
where the collective index $\alpha $ is a set of quantum numbers specifying
the solution.

To evaluate the mode sum in formula (\ref{mswf}) we need the form of the
eigenfunctions $\varphi _{\alpha }(x)$ (for a recent discussion of
eigenmodes in four Rindler sectors and relations between them see, for
instance, \cite{Gerl99}). For the geometry under consideration the metric
and boundary conditions are static and translational invariant in the
hyperplane parallel to the plates. It follows from here that the
corresponding part of the eigenfunctions can be taken in the standard plane
wave form:
\begin{equation}
\varphi _{\alpha }=C\phi (\xi )\exp \left[ i\left( \mathbf{kx}-\omega \tau
\right) \right] ,\quad \alpha =(\mathbf{k},\omega ),  \label{wavesracture}
\end{equation}%
with the wave vector $\mathbf{k}=(k_{2},\ldots ,k_{D})$. The frequency $%
\omega $ in Eq. (\ref{wavesracture}) corresponds to the dimensionless
coordinate $\tau $ and hence is dimensionless. The proper time $\tau _{g}$
and the frequency $\omega _{g}$ measured by a uniformly accelerated observer
with the proper acceleration $g$ and world line $(x^{1})^{2}-t^{2}=g^{-2}$
are related to $\tau $ and $\omega $ by the formulas $\tau _{g}=\tau /g$, $%
\omega _{g}=\omega g$ (the features of the measurements for time, frequency,
and length relative to a Rindler frame as compared to a Minkowski frame are
discussed in Ref. \cite{Gerl03}). The equation for $\phi (\xi )$ is obtained
from field equation (\ref{fieldeq}) on background of metric (\ref{metric})
and has the form
\begin{equation}
\xi ^{2}\phi ^{\prime \prime }(\xi )+\xi \phi ^{\prime }(\xi )+\left( \omega
^{2}-\lambda ^{2}\xi ^{2}\right) \phi (\xi )=0,  \label{fiequ}
\end{equation}%
where the prime denotes a differentiation with respect to the argument of
the function,
\begin{equation}
\lambda =\sqrt{k^{2}+m^{2}},  \label{lambda}
\end{equation}%
and $k=|\mathbf{k}|$. In the region between the plates the linearly
independent solutions to equation (\ref{fiequ}) are the Bessel modified
functions $I_{i\omega }(k\xi )$ and $K_{i\omega }(k\xi )$. The solution
satisfying boundary condition (\ref{Dboundcond}) on the plate $\xi =b$ has
the form
\begin{equation}
Z_{i\omega }^{(b)}(\lambda \xi ,\lambda b)=\bar{I}_{i\omega }^{(b)}(\lambda
b)K_{i\omega }(\lambda \xi )-\bar{K}_{i\omega }^{(b)}(\lambda b)I_{i\omega
}(\lambda \xi ).  \label{Deigfunc}
\end{equation}%
Here and below for a given function $f(z)$ we use the notations
\begin{equation}
\bar{f}^{(j)}(z)=A_{j}f(z)+\frac{B_{j}}{j}zf^{\prime }(z),\quad j=a,b.
\label{fbarnot}
\end{equation}%
Note that function (\ref{Deigfunc}) is real, $Z_{i\omega }^{(b)}(\lambda \xi
,\lambda b)=Z_{-i\omega }^{(b)}(\lambda \xi ,\lambda b)$. From the boundary
condition on the plate $\xi =a$ we find that the possible values for $\omega
$ are roots to the equation
\begin{equation}
Z_{i\omega }(\lambda a,\lambda b)=0,  \label{Deigfreq}
\end{equation}%
with the notation
\begin{equation}
Z_{\omega }(u,v)=\bar{I}_{\omega }^{(b)}(v)\bar{K}_{\omega }^{(a)}(u)-\bar{K}%
_{\omega }^{(b)}(v)\bar{I}_{\omega }^{(a)}(u).  \label{Zomega}
\end{equation}%
For a fixed $\lambda $, the equation (\ref{Deigfreq}) has an infinite set of
real solutions with respect to $\omega $. We will denote them by $\omega
_{n}=\omega _{n}(\lambda a,\lambda b)$, $\omega _{n}>0$, $n=1,2,\ldots $,
and will assume that they are arranged in the ascending order $\omega
_{n}<\omega _{n+1}$. In addition to the real zeros, in dependence of the
values of the ratios $A_{j}/B_{j}$, equation (\ref{Deigfreq}) can have a
finite set of purely imaginary solutions. The presence of such solutions
leads to the modes with an imaginary frequency and, hence, to the unstable
vacuum. In the consideration below we will assume the values of the
coefficients in Eq. (\ref{Dboundcond}) for which the imaginary solutions are
absent and the vacuum is stable.

The coefficient $C$ in formula (\ref{wavesracture}) is determined from the
standard Klein-Gordon orthonormality condition for the eigenfunctions which
for metric (\ref{metric}) takes the form
\begin{equation}
\int d\mathbf{x}\int_{a}^{b}\frac{d\xi }{\xi }\varphi _{\alpha }%
\overleftrightarrow{\partial }_{\tau }\varphi _{\alpha ^{\prime }}^{\ast
}=i\delta _{\alpha \alpha ^{\prime }}.  \label{normcond}
\end{equation}%
The $\xi $-integral on the left of this formula is evaluated using the
integration formula
\begin{equation}
\int_{a}^{b}\frac{d\xi }{\xi }\phi _{1\omega }(\xi )\phi _{2v}(\xi )=\xi
\left. \frac{\phi _{1\omega }(\xi )\phi _{2\upsilon }^{\prime }(\xi )-\phi
_{2\nu }(\xi )\phi _{1\omega }^{\prime }(\xi )}{\omega ^{2}-\upsilon ^{2}}%
\right\vert _{a}^{b},  \label{intformula}
\end{equation}%
valid for any two solutions $\phi _{l\omega }(\xi )$, $l=1,2$, to equation (%
\ref{fiequ}). Taking into account boundary condition (\ref{Dboundcond}),
from Eq. (\ref{normcond}) for the normalization coefficient one finds
\begin{equation}
C^{2}=\left. \frac{\left( 2\pi \right) ^{1-D}\bar{I}_{i\omega
}^{(a)}(\lambda a)}{\bar{I}_{i\omega }^{(b)}(\lambda b)\frac{\partial }{%
\partial \omega }Z_{i\omega }(\lambda a,\lambda b)}\right\vert _{\omega
=\omega _{n}}.  \label{Dnormc}
\end{equation}%
Now substituting the eigenfunctions
\begin{equation}
\varphi _{\alpha }(x)=CZ_{i\omega _{n}}^{(b)}(\lambda \xi ,\lambda b)\exp
\left[ i\left( \mathbf{kx}-\omega _{n}\tau \right) \right]  \label{Dsol2}
\end{equation}%
into the mode sum formula (\ref{mswf}), for the positive frequency Wightman
function one finds
\begin{eqnarray}
G^{+}(x,x^{\prime }) &=&\int \frac{d\mathbf{k}e^{i\mathbf{k}(\mathbf{x}-%
\mathbf{x}^{\prime })}}{(2\pi )^{D-1}}\sum_{n=1}^{\infty }\frac{\bar{I}%
_{i\omega }^{(a)}(\lambda a)e^{-i\omega (\tau -\tau ^{\prime })}}{\bar{I}%
_{i\omega }^{(b)}(\lambda b)\frac{\partial }{\partial \omega }Z_{i\omega
}(\lambda a,\lambda b)}  \notag \\
&&\times \left. Z_{i\omega }^{(b)}(\lambda \xi ,\lambda b)Z_{i\omega
}^{(b)}(\lambda \xi ^{\prime },\lambda b)\right\vert _{\omega =\omega _{n}}.
\label{Wigh1}
\end{eqnarray}%
As the expressions for the eigenfrequencies $\omega _{n}$ (as functions on $%
\lambda a$, $\lambda b$, $A_{j}/B_{j}$) are not explicitly known, the form (%
\ref{Wigh1}) of the Wightman function is inconvenient. For the further
evoluation of this VEV we can apply to the sum over $n$ the summation
formula (\ref{Dsumformula}) derived in Appendix \ref{section:App1} on the
base of the generalized Abel-Plana formula. As a function $F(z)$ in formula (%
\ref{Dsumformula}) let us choose
\begin{equation}
F(z)=\frac{Z_{iz}^{(b)}(\lambda \xi ,\lambda b)Z_{iz}^{(b)}(\lambda \xi
^{\prime },\lambda b)}{\bar{I}_{iz}^{(b)}(\lambda b)\bar{I}%
_{-iz}^{(b)}(\lambda b)}e^{-iz(\tau -\tau ^{\prime })}.  \label{FtoAPF}
\end{equation}%
Condition (\ref{condforAPF2pl}) for this function is satisfied if $%
a^{2}e^{|\tau -\tau ^{\prime }|}<\xi \xi ^{\prime }$. In particular, this is
the case in the coincidence limit $\tau =\tau ^{\prime }$ in the region
under consideration: $\xi ,\xi ^{\prime }>a$. By using formula (\ref%
{Dsumformula}), for the Wightman function one obtains the expression%
\begin{eqnarray}
G^{+}(x,x^{\prime }) &=&G^{+}(x,x^{\prime };b)-\int \frac{d\mathbf{k\,}e^{i%
\mathbf{k}(\mathbf{x}-\mathbf{x}^{\prime })}}{\pi (2\pi )^{D-1}}%
\int_{0}^{\infty }d\omega \,\Omega _{b\omega }(\lambda a,\lambda b)  \notag
\\
&&\times Z_{\omega }^{(b)}(\lambda \xi ,\lambda b)Z_{\omega }^{(b)}(\lambda
\xi ^{\prime },\lambda b)\cosh [\omega (\tau -\tau ^{\prime })],
\label{Wigh3}
\end{eqnarray}%
where we have introduced the notation%
\begin{equation}
\Omega _{b\omega }(\lambda a,\lambda b)=\frac{\bar{I}_{\omega
}^{(a)}(\lambda a)}{\bar{I}_{\omega }^{(b)}(\lambda b)Z_{\omega }(\lambda
a,\lambda b)}.  \label{Omega2}
\end{equation}%
In Eq. (\ref{Wigh3})
\begin{eqnarray}
G^{+}(x,x^{\prime };b) &=&\int \frac{d\mathbf{k}e^{i\mathbf{k}(\mathbf{x}-%
\mathbf{x}^{\prime })}}{\pi ^{2}(2\pi )^{D-1}}\int_{0}^{\infty }d\omega
\sinh (\pi \omega )  \notag \\
&&\times e^{-i\omega (\tau -\tau ^{\prime })}\frac{Z_{i\omega
}^{(b)}(\lambda \xi ,\lambda b)Z_{i\omega }^{(b)}(\lambda \xi ^{\prime
},\lambda b)}{\bar{I}_{i\omega }^{(b)}(\lambda b)\bar{I}_{-i\omega
}^{(b)}(\lambda b)},  \label{Wigh1pl}
\end{eqnarray}%
is the Wightman function in the region $\xi <b$ for a single plate at $\xi
=b $. This function is investigated in Ref. \cite{Saha02} and can be
presented in the form
\begin{equation}
G^{+}(x,x^{\prime };b)=G_{R}^{+}(x,x^{\prime })+\left\langle \varphi
(x)\varphi (x^{\prime })\right\rangle ^{(b)},  \label{G+2}
\end{equation}%
where $G_{R}^{+}(x,x^{\prime })$ is the Wightman function for the right
Rindler wedge without boundaries and the part%
\begin{eqnarray}
\left\langle \varphi (x)\varphi (x^{\prime })\right\rangle ^{(b)} &=&-\int
\frac{d\mathbf{k}e^{i\mathbf{k}(\mathbf{x}-\mathbf{x}^{\prime })}}{\pi (2\pi
)^{D-1}}\int_{0}^{\infty }d\omega \frac{\bar{K}_{\omega }^{(b)}(\lambda b)}{%
\bar{I}_{\omega }^{(b)}(\lambda b)}  \notag \\
&&\times I_{\omega }(\lambda \xi )I_{\omega }(\lambda \xi ^{\prime })\cosh
[\omega (\tau -\tau ^{\prime })]  \label{phi212}
\end{eqnarray}%
is induced in the region $\xi <b$ by the presence of the plate at $\xi =b$.
Note that the representation (\ref{G+2}) with (\ref{phi212}) is valid under
the assumption $\xi \xi ^{\prime }<b^{2}e^{|\tau -\tau ^{\prime }|}$. Hence,
the application of the summation formula based on the generalized Abel-Plana
formula allowed us (i) to escape the necessity to know the explicit
expressions for the eigenfrequencies $\omega _{n}$, (ii) to extract from the
VEVs the purely Rindler and single plate parts, (iii) to present the
remained part in terms of integrals with the exponential convergence in the
coincidence limit.

By using the identity%
\begin{eqnarray}
&&\frac{\bar{K}_{\omega }^{(b)}(\lambda b)}{\bar{I}_{\omega }^{(b)}(\lambda
b)}I_{\omega }(\lambda \xi )I_{\omega }(\lambda \xi ^{\prime })=\frac{\bar{I}%
_{\omega }^{(a)}(\lambda a)}{\bar{K}_{\omega }^{(a)}(\lambda a)}K_{\omega
}(\lambda \xi )K_{\omega }(\lambda \xi ^{\prime })  \notag \\
&&+\sum_{j=a,b}n^{(j)}\Omega _{j\omega }(\lambda a,\lambda b)Z_{\omega
}^{(j)}(\lambda \xi ,\lambda j)Z_{\omega }^{(j)}(\lambda \xi ^{\prime
},\lambda j),  \label{ident1}
\end{eqnarray}%
with $n^{(a)}=1$, $n^{(b)}=-1$, the Wightman function can be also presented
in the form%
\begin{eqnarray}
G^{+}(x,x^{\prime }) &=&G^{+}(x,x^{\prime };a)-\int \frac{d\mathbf{k\,}e^{i%
\mathbf{k}(\mathbf{x}-\mathbf{x}^{\prime })}}{\pi (2\pi )^{D-1}}%
\int_{0}^{\infty }d\omega \,\Omega _{a\omega }(\lambda a,\lambda b)  \notag
\\
&&\times Z_{\omega }^{(a)}(\lambda \xi ,\lambda a)Z_{\omega }^{(a)}(\lambda
\xi ^{\prime },\lambda a)\cosh [\omega (\tau -\tau ^{\prime })].
\label{Wigh31}
\end{eqnarray}%
In this formula%
\begin{equation}
G^{+}(x,x^{\prime };a)=G_{R}^{+}(x,x^{\prime })+\left\langle \varphi
(x)\varphi (x^{\prime })\right\rangle ^{(a)}  \label{G+1}
\end{equation}%
is the Wightman function in the region $\xi >a$ for a single plate at $\xi
=a $, and
\begin{eqnarray}
\left\langle \varphi (x)\varphi (x^{\prime })\right\rangle ^{(a)} &=&-\int
\frac{d\mathbf{k}e^{i\mathbf{k}(\mathbf{x}-\mathbf{x}^{\prime })}}{\pi (2\pi
)^{D-1}}\int_{0}^{\infty }d\omega \frac{\bar{I}_{\omega }^{(a)}(\lambda a)}{%
\bar{K}_{\omega }^{(a)}(\lambda a)}  \notag \\
&&\times K_{\omega }(\lambda \xi )K_{\omega }(\lambda \xi ^{\prime })\cosh
[\omega (\tau -\tau ^{\prime })].  \label{phi211}
\end{eqnarray}%
In Eq. (\ref{ident1}) we use the notations%
\begin{eqnarray}
Z_{\omega }^{(a)}(\lambda \xi ,\lambda a) &=&\bar{I}_{\omega }^{(a)}(\lambda
a)K_{\omega }(\lambda \xi )-\bar{K}_{\omega }^{(a)}(\lambda a)I_{\omega
}(\lambda \xi ),  \label{Zom1} \\
\Omega _{a\omega }(\lambda a,\lambda b) &=&\frac{\bar{K}_{\omega
}^{(b)}(\lambda b)}{\bar{K}_{\omega }^{(a)}(\lambda a)Z_{\omega }(\lambda
a,\lambda b)}.  \label{Oma}
\end{eqnarray}%
Two representations of the Wightman function, Eqs. (\ref{Wigh3}) and (\ref%
{Wigh31}), are obtained from each other by the replacements%
\begin{equation}
a\rightleftarrows b,\quad I_{\omega }\rightleftarrows K_{\omega }.
\label{replacement}
\end{equation}%
In the coincidence limit the second term on the right of formula (\ref{Wigh3}%
) is finite on the plate $\xi =b$ and diverges on the plate at $\xi =a$,
whereas the second term on the right of Eq. (\ref{Wigh31}) is finite on the
plate $\xi =a$ and is divergent for $\xi =b$. Consequently, the form (\ref%
{Wigh3}) [(\ref{Wigh31})] is convenient for the investigations of the VEVs
near the plate $\xi =b$ ($\xi =a$). Note that in the formulas given above
the integration over the angular part of the vector $\mathbf{k}$ can be done
with the help of the formula%
\begin{equation}
\int d{\mathbf{k}}\,\frac{e^{i{\mathbf{kx}}}F(k)}{(2\pi )^{\frac{D-1}{2}}}%
=\int_{0}^{\infty }dk\,k^{D-2}F(k)\frac{J_{(D-3)/2}(k|{\mathbf{x}}|)}{(k|{%
\mathbf{x}}|)^{(D-3)/2}},  \label{intformwf}
\end{equation}%
for a given function $F(k)$, and $J_{\nu }(z)$ is the Bessel function. In
this section we have considered the positive frequency Whightman function.
By the same method any other two-point function (Hadamard function,
Feynman's Green function, etc.) can be evaluated.

\section{Casimir densities}

\label{sec:VEVEMT}

\subsection{VEV for the field square}

In this section we will consider the VEVs of the field square and the
energy-momentum tensor in the region between the plates. As the
corresponding quantities for a single plate are investigated in Ref. \cite%
{Saha02}, here we will be concentrated on the parts induced by the presence
of the second plate. In the coincidence limit from the formulas for the
Wightman function one obtains two equivalent forms for the VEV\ of the field
square:%
\begin{eqnarray}
\left\langle 0\left\vert \varphi ^{2}\right\vert 0\right\rangle
&=&\left\langle 0_{R}\left\vert \varphi ^{2}\right\vert 0_{R}\right\rangle
+\left\langle \varphi ^{2}\right\rangle ^{(j)}  \notag \\
&&-A_{D}\int_{0}^{\infty }dk\,k^{D-2}\int_{0}^{\infty }d\omega \,\Omega
_{j\omega }(\lambda a,\lambda b)Z_{\omega }^{(j)2}(\lambda \xi ,\lambda j),
\label{phi2sq1}
\end{eqnarray}%
corresponding to $j=a$ and $j=b$, and $\left. |0_{R}\right\rangle $ is the
amplitude for the Fulling-Rindler vacuum without boundaries,%
\begin{equation}
A_{D}=\frac{1}{2^{D-2}\pi ^{\frac{D+1}{2}}\Gamma \left( \frac{D-1}{2}\right)
}.  \label{Ad}
\end{equation}%
In Eq. (\ref{phi2sq1}) the part $\left\langle \varphi ^{2}\right\rangle
^{(j)}$ is induced by a single plate at $\xi =j$ when the second plate is
absent. From (\ref{phi212}), (\ref{phi211}) for this part one has \cite%
{Saha02}
\begin{subequations}
\label{phi21plgen}
\begin{eqnarray}
\left\langle \varphi ^{2}\right\rangle ^{(a)} &=&-A_{D}\int_{0}^{\infty
}dk\,k^{D-2}\int_{0}^{\infty }d\omega \frac{\bar{I}_{\omega }^{(a)}(\lambda
a)}{\bar{K}_{\omega }^{(a)}(\lambda a)}K_{\omega }^{2}(\lambda \xi ),
\label{phi21pl} \\
\left\langle \varphi ^{2}\right\rangle ^{(b)} &=&-A_{D}\int_{0}^{\infty
}dk\,k^{D-2}\int_{0}^{\infty }d\omega \frac{\bar{K}_{\omega }^{(b)}(\lambda
b)}{\bar{I}_{\omega }^{(b)}(\lambda b)}I_{\omega }^{2}(\lambda \xi ).
\label{phi21plb}
\end{eqnarray}%
The last term on the right of formula (\ref{phi2sq1}) is finite on the plate
at $\xi =j$ and diverges for the points on the other plate.

Extracting the contribution from the second plate, we can write the
expression (\ref{phi2sq1}) for the vacuum expectation value in the symmetric
form
\end{subequations}
\begin{equation}
\left\langle 0\left\vert \varphi ^{2}\right\vert 0\right\rangle
=\left\langle 0_{R}\left\vert \varphi ^{2}\right\vert
0_{R}\right\rangle +\sum_{j=a,b}\left\langle \varphi
^{2}\right\rangle ^{(j)}+\left\langle \varphi ^{2}\right\rangle
^{(ab)},  \label{phi2sq2n}
\end{equation}%
with the 'interference' part%
\begin{eqnarray}
\left\langle \varphi ^{2}\right\rangle ^{(ab)} &=&-A_{D}\int_{0}^{\infty
}dk\,k^{D-2}\int_{0}^{\infty }d\omega \bar{I}_{\omega }^{(a)}(\lambda a)
\notag \\
&&\times \left[ \frac{Z_{\omega }^{(b)2}(\lambda \xi ,\lambda b)}{\bar{I}%
_{\omega }^{(b)}(\lambda b)Z_{\omega }(\lambda a,\lambda b)}-\frac{K_{\omega
}^{2}(\lambda \xi )}{\bar{K}_{\omega }^{(a)}(\lambda a)}\right] .
\label{phi2int}
\end{eqnarray}%
An equivalent form for this part is obtained with the replacements (\ref%
{replacement}) in the subintegrand. The 'interference' term (\ref{phi2int})
is finite for all values of $\xi $ in the range $a\leq \xi \leq b$,
including the points on the boundaries. The well-known surface divergences
are contained in the single plate parts only. To find the corresponding
asymptotic behaviour we note that for the points near the boundaries the
main contribution into the $\omega $-integral comes from large values of $%
\omega $ and we can use the uniform asymptotic expansions for the modified
Bessel functions for large values of the order (see, for instance, \cite%
{Abramowitz}). Introducing a new integration variable $x=k/\omega $ and
replacing the modified Bessel functions by their uniform asymptotic
expansions, in the limit $\xi \rightarrow j$ to the leading order one obtains%
\begin{equation}
\left\langle \varphi ^{2}\right\rangle ^{(j)}\approx \frac{k_{j}\Gamma
\left( \frac{D-1}{2}\right) }{(4\pi )^{\frac{D+1}{2}}|\xi -j|^{D-1}},
\label{phi2asnear}
\end{equation}%
where%
\begin{equation}
k_{j}=1-2\delta _{B_{j}0}.  \label{kj}
\end{equation}%
This term has different signs for Dirichlet and non-Dirichlet boundary
conditions and is the same as that for a plate on the Minkowski bulk with $%
|\xi -j|$ being the distance from the plate.

In the limit $a\rightarrow b$ with the fixed values of the coefficients in
the boundary conditions and the mass, the 'interference' part (\ref{phi2int}%
) is divergent and for small values of $b/a-1$ the main contribution comes
from large values of $\omega $. Again, introducing an integration variable $%
x=k/\omega $ and replacing the modified Bessel functions by their uniform
asymptotic expansions, to the leading order one obtains%
\begin{equation}
\langle \varphi ^{2}\rangle ^{(ab)}\approx \frac{(4\pi )^{-\frac{D}{2}}}{%
\Gamma \left( \frac{D}{2}\right) }\int_{0}^{\infty }dy\,y^{D-1}\frac{%
k_{a}e^{2y(a-\xi )}+k_{b}e^{2y(\xi -b)}+2}{k_{a}k_{b}e^{2y(b-a)}-1}.
\label{phi2closeab}
\end{equation}

Large values of the proper accelerations for the plates correspond to the
limit $a,b\rightarrow 0$. In this limit the plates are close to the Rindler
horizon. From formulas (\ref{phi21pl}), (\ref{phi21plb}), (\ref{phi2int}) we
see that for fixed values of the ratios $a/b$, $\xi /b$, both single plate
and 'interference' parts behave as $b^{1-D}$ in the limit $b\rightarrow 0$.
In the limit $a\rightarrow 0$ for fixed values $\xi $ and $b$, the left
plate tends to the Rindler horizon for a fixed world line of the right
plate. The main contribution into the $\omega $-integral in Eq. (\ref%
{phi2int}) comes from small values $\omega $, $\omega \lesssim 1/\ln
(2/\lambda a)$. Using the formulas for the Bessel modified functions for
small arguments, it can be seen that the 'interference' part (\ref{phi2int})
vanishes as $\ln ^{-2}(2b/a)$.

Now we turn to the limit of small accelerations of the plates: $%
a,b\rightarrow \infty $ with fixed values $b-a$, $B_{j}/A_{j}$, and $m$. In
this case the main contribution comes from large values of $\omega $. Using
the uniform asymptotic formulas for the Bessel modified functions, the
following formula is obtained for the single plate parts:%
\begin{equation}
\left\langle \varphi ^{2}\right\rangle ^{(j)}\approx -\frac{(4\pi )^{-\frac{D%
}{2}}}{\Gamma \left( \frac{D}{2}\right) }\int_{m}^{\infty }dy\,\left(
y^{2}-m^{2}\right) ^{\frac{D}{2}-1}\frac{e^{-2y|\xi -j|}}{c_{j}(y)},
\label{phi21plclose}
\end{equation}%
with the notation%
\begin{equation}
c_{j}(y)=\frac{A_{j}-n^{(j)}B_{j}y}{A_{j}+n^{(j)}B_{j}y},\quad
n^{(a)}=1,\quad n^{(b)}=-1.  \label{cjyn}
\end{equation}%
Similarly, for the 'interference' term we find
\begin{equation}
\langle \varphi ^{2}\rangle ^{(ab)}\approx \frac{(4\pi )^{-\frac{D}{2}}}{%
\Gamma \left( \frac{D}{2}\right) }\int_{m}^{\infty }dy\,\frac{\left(
y^{2}-m^{2}\right) ^{\frac{D}{2}-1}}{c_{a}(y)c_{b}(y)e^{2y(b-a)}-1}\left[
2-\sum_{j=a,b}\frac{e^{-2y|\xi -j|}}{c_{j}(y)}\right] .  \label{phi2close}
\end{equation}%
Formulae (\ref{phi21plclose}) and (\ref{phi2close}) coincide with the
corresponding expressions for the geometry of two parallel plates on the
Minkowski bulk. In this limit, $\xi $ corresponds to the Cartesian
coordinate perpendicular to the plates which are located at $\xi =a$ and $%
\xi =b$.

For large values of the mass, $ma\gg 1$, we introduce in (\ref{phi21pl}) and
(\ref{phi21plb}) a new integration variable $y=\lambda /m$. The main
contribution into the $\omega $-integral comes from the values $\omega \sim
\sqrt{ma}$. By using the uniform asymptotic expansions for the Bessel
modified functions for large values of the order and further expanding over $%
\omega /ma$, for the single plate parts to the leading order one finds%
\begin{equation}
\left\langle \varphi ^{2}\right\rangle ^{(j)}\approx -\frac{m^{\frac{D}{2}%
-1}e^{-2m|\xi -j|}\sqrt{j/\xi }}{2(4\pi )^{\frac{D}{2}}c_{j}(m)|\xi -j|^{%
\frac{D}{2}}},  \label{phi2largemass}
\end{equation}%
for $j=a,b$. By the similar way, for the 'interference' part we obtain the
formula%
\begin{equation}
\langle \varphi ^{2}\rangle ^{(ab)}\approx \frac{m^{\frac{D}{2}-1}e^{2m(a-b)}%
\sqrt{ab}}{(4\pi )^{\frac{D}{2}}\xi c_{a}(m)c_{b}(m)(b-a)^{\frac{D}{2}}}.
\label{phi2largemassab}
\end{equation}%
As we could expect, the both single plate and 'interference' parts are
exponentially suppressed for large values of the mass.

\subsection{VEV of the energy-momentum tensor}

By using the field equation it can be seen that the expression for the
energy-momentum tensor of the scalar field under consideration can be
presented in the form
\begin{equation}
T_{ik}=\nabla _{i}\varphi \nabla _{k}\varphi +\left[ \left( \zeta -\frac{1}{4%
}\right) g_{ik}\nabla _{l}\nabla ^{l}-\zeta \nabla _{i}\nabla _{k}\right]
\varphi ^{2},  \label{EMT2}
\end{equation}%
and the corresponding trace is equal to
\begin{equation}
T_{i}^{i}=D(\zeta -\zeta _{c})\nabla _{i}\nabla ^{i}\varphi
^{2}+m^{2}\varphi ^{2}.  \label{trace}
\end{equation}%
By virtue of Eq. (\ref{EMT2}), the VEV of the energy-momentum tensor is
expressed in terms of the Wightman function as
\begin{equation}
\langle 0\left\vert T_{ik}(x)\right\vert 0\rangle =\lim_{x^{\prime
}\rightarrow x}\nabla _{i}\nabla _{k}^{\prime }G^{+}(x,x^{\prime })+\left[
\left( \zeta -\frac{1}{4}\right) g_{ik}\nabla _{l}\nabla ^{l}-\zeta \nabla
_{i}\nabla _{k}\right] \langle 0\left\vert \varphi ^{2}(x)\right\vert
0\rangle .  \label{vevemtW}
\end{equation}%
Making use the formulas for the Wightman function and the field square, one
obtains two equivalent forms, corresponding to $j=a$ and $j=b$ (no summation
over $i$):
\begin{eqnarray}
\langle 0|T_{i}^{k}|0\rangle &=&\langle 0_{R}|T_{i}^{k}|0_{R}\rangle
+\langle T_{i}^{k}\rangle ^{(j)}-A_{D}\delta _{i}^{k}\int dk\,k^{D-2}  \notag
\\
&&\times \lambda ^{2}\int_{0}^{\infty }d\omega \,\Omega _{j\omega }(\lambda
a,\lambda b)F^{(i)}\left[ Z_{\omega }^{(j)}(\lambda \xi ,\lambda j)\right] .
\label{Tik1}
\end{eqnarray}%
In this formula,
\begin{equation}
\langle 0_{R}|T_{i}^{k}|0_{R}\rangle =\delta _{i}^{k}\frac{A_{D}}{\pi }%
\int_{0}^{\infty }dkk^{D-2}\lambda ^{2}\int_{0}^{\infty }d\omega \sinh \pi
\omega \,f^{(i)}[K_{i\omega }(\lambda \xi )]  \label{DFR}
\end{equation}%
is the corresponding VEV for the Fulling--Rindler vacuum without boundaries,
and the terms (no summation over $i$)
\begin{subequations}
\label{D1plateboundgen}
\begin{eqnarray}
\langle T_{i}^{k}\rangle ^{(a)} &=&-A_{D}\delta _{i}^{k}\int_{0}^{\infty
}dkk^{D-2}\lambda ^{2}\int_{0}^{\infty }d\omega \frac{\bar{I}_{\omega
}^{(a)}(\lambda a)}{\bar{K}_{\omega }^{(a)}(\lambda a)}F^{(i)}[K_{\omega
}(\lambda \xi )],  \label{D1platebound} \\
\langle T_{i}^{k}\rangle ^{(b)} &=&-A_{D}\delta _{i}^{k}\int_{0}^{\infty
}dkk^{D-2}\lambda ^{2}\int_{0}^{\infty }d\omega \frac{\bar{K}_{\omega
}^{(b)}(\lambda b)}{\bar{I}_{\omega }^{(b)}(\lambda b)}F^{(i)}[I_{\omega
}(\lambda \xi )],  \label{D1plateboundb}
\end{eqnarray}%
are induced by the presence of a single plane boundaries located at $\xi =a$
and $\xi =b$ in the regions $\xi >a$ and $\xi <b$ respectively. In formulas (%
\ref{Tik1}), (\ref{D1platebound}), (\ref{D1plateboundb}) for a given
function $g(z)$ we use the notations
\end{subequations}
\begin{eqnarray}
F^{(0)}[g(z)] &=&\left( \frac{1}{2}-2\zeta \right) \left[ \left( \frac{dg(z)%
}{dz}\right) ^{2}+\left( 1+\frac{\omega ^{2}}{z^{2}}\right) g^{2}(z)\right] +%
\frac{\zeta }{z}\frac{d}{dz}g^{2}(z)-\frac{\omega ^{2}}{z^{2}}g^{2}(z),
\label{f0} \\
F^{(1)}[g(z)] &=&-\frac{1}{2}\left( \frac{dg(z)}{dz}\right) ^{2}-\frac{\zeta
}{z}\frac{d}{dz}g^{2}(z)+\frac{1}{2}\left( 1+\frac{\omega ^{2}}{z^{2}}%
\right) g^{2}(z),  \label{f1} \\
F^{(i)}[g(z)] &=&\left( \frac{1}{2}-2\zeta \right) \left[ \left( \frac{dg(z)%
}{dz}\right) ^{2}+\left( 1+\frac{\omega ^{2}}{z^{2}}\right) g^{2}(z)\right] -%
\frac{g^{2}(z)}{D-1}\frac{k^{2}}{\lambda ^{2}},  \label{f23}
\end{eqnarray}%
where $i=2,\ldots ,D$ and the indices 0,1 correspond to the coordinates $%
\tau $, $\xi $, respectively. For the last term on the right of Eq. (\ref%
{Tik1}) we have to substitute $g(z)=Z_{\omega }^{(j)}(z,\lambda j)$. The
expressions for the functions $f^{(i)}[g(z)]$ in (\ref{DFR}) are obtained
from the corresponding expressions for $F^{(i)}[g(z)]$ by the replacement $%
\omega \rightarrow i\omega $. It can be easily seen that for a conformally
coupled massless scalar the energy-momentum tensor is traceless.

The purely Fulling-Rindler part (\ref{DFR}) of the energy-momentum tensor is
investigated in a large number of papers (see, for instance, references
given in \cite{Avag02}). The most general case of a massive scalar field in
an arbitrary number of spacetime dimensions has been considered in Ref. \cite%
{Hill} for conformally and minimally coupled cases and in Ref. \cite{Saha02}
for general values of the curvature coupling parameter. For a massless
scalar the VEV for the Rindler part without boundaries can be presented in
the form
\begin{eqnarray}
\langle T_{i}^{k}\rangle _{\mathrm{sub}}^{(R)} &=&\langle
0_{R}|T_{i}^{k}|0_{R}\rangle -\langle 0_{M}|T_{i}^{k}|0_{M}\rangle  \notag \\
&=&-\frac{2\delta _{i}^{k}\xi ^{-D-1}}{(4\pi )^{\frac{D}{2}}\Gamma \left(
\frac{D}{2}\right) }\int_{0}^{\infty }\frac{\omega ^{D}g^{(i)}(\omega
)d\omega }{e^{2\pi \omega }+(-1)^{D}}\,,  \label{subRindm0}
\end{eqnarray}%
where the expressions for the functions $g^{(i)}(\omega )$ are presented in
Ref. \cite{Saha02}, and $\left. |0_{M}\right\rangle $ is the amplitude for
the Minkowski vacuum without boundaries. Expression (\ref{subRindm0})
corresponds to the absence from the vacuum of thermal distribution with
standard temperature $T=(2\pi \xi )^{-1}$. In general, the corresponding
spectrum has non-Planckian form: the density of states factor is not
proportional to $\omega ^{D-1}d\omega $. The spectrum takes the Planckian
form for conformally coupled scalars in $D=1,2,3$ with $g^{(0)}(\omega
)=-Dg^{(i)}(\omega )=1$, $i=1,2,\ldots D$. It is of interest to note that
for even values of spatial dimension the distribution is Fermi-Dirac type
(see also \cite{Taga85,Oogu86}). For the massive scalar the energy spectrum
is not strictly thermal and the corresponding quantities do not coincide
with ones for the Minkowski thermal bath.

The boundary induced quantities (\ref{D1platebound}), (\ref{D1plateboundb})
are investigated in Ref. \cite{Candelas} for a conformally coupled $D=3$
massless Dirichlet scalar in the region on the right from a single plate and
in Ref. \cite{Saha02} for a massive scalar with general curvature coupling
and Robin boundary condition in an arbitrary number of dimensions in both
regions. The single boundary parts diverge at the plates surfaces $\xi =j$, $%
j=a,b$. Near the plates the leading terms of the corresponding asymptotic
expansions have the form (no summation over $i$)
\begin{equation}
\langle T_{i}^{i}\rangle ^{(j)}\approx \frac{Dj\langle T_{1}^{1}\rangle
^{(j)}}{(D-1)(j-\xi )}\approx \frac{D(\zeta _{c}-\zeta )\Gamma \left( \frac{%
D+1}{2}\right) }{2^{D}\pi ^{\frac{D+1}{2}}|\xi -j|^{D+1}}k_{j},
\label{1basymppe}
\end{equation}%
with $i=0,2,\ldots ,D$, and $k_{j}$ is defined by Eq. (\ref{kj}). These
leading terms vanish for a conformally coupled scalar and coincide with the
corresponding quantities for a plane boundary in the Minkowski vacuum.

Now let us present the VEV (\ref{Tik1}) in the form%
\begin{equation}
\langle 0|T_{i}^{k}|0\rangle =\langle 0_{R}|T_{i}^{k}|0_{R}\rangle
+\sum_{j=a,b}\langle T_{i}^{k}\rangle ^{(j)}+\langle T_{i}^{k}\rangle
^{(ab)},  \label{Tikdecomp}
\end{equation}%
where (no summation over $i$)
\begin{eqnarray}
\langle T_{i}^{k}\rangle ^{(ab)} &=&-A_{D}\delta _{i}^{k}\int_{0}^{\infty
}dk\,k^{D-2}\lambda ^{2}\int_{0}^{\infty }d\omega \bar{I}_{\omega
}^{(a)}(\lambda a)  \notag \\
&&\times \left[ \frac{F^{(i)}[Z_{\omega }^{(b)}(\lambda \xi ,\lambda b)]}{%
\bar{I}_{\omega }^{(b)}(\lambda b)Z_{\omega }(\lambda a,\lambda b)}-\frac{%
F^{(i)}[K_{\omega }(\lambda \xi )]}{\bar{K}_{\omega }^{(a)}(\lambda a)}%
\right]  \label{intterm1}
\end{eqnarray}%
is the 'interference' term. The surface divergences are contained in the
single boundary parts and this term is finite for all values $a\leq \xi \leq
b$. An equivalent formula for $\langle T_{i}^{k}\rangle ^{(ab)}$ is obtained
from Eq. (\ref{intterm1}) by replacements (\ref{replacement}).

Both single plate and 'interference' parts separately satisfy the standard
continuity equation for the energy-momentum tensor, which for the geometry
under consideration takes the form
\begin{equation}
\frac{d(\xi \left\langle T_{1}^{1}\right\rangle )}{d\xi }=\left\langle
T_{0}^{0}\right\rangle .  \label{rel}
\end{equation}%
For a conformally coupled massless scalar field the both parts are traceless
and we have an additional relation $\left\langle T_{i}^{i}\right\rangle =0$.

In the limit $a\rightarrow b$ expression (\ref{intterm1}) is divergent and
for small values of $b/a-1$ the main contribution comes from the large
values of $\omega $. Introducing a new integration variable $x=k/\omega $
and replacing Bessel modified functions by their uniform asymptotic
expansions for large values of the order, at the leading order one receives%
\begin{eqnarray}
\langle T_{i}^{i}\rangle ^{(ab)} &\approx &-\frac{(4\pi )^{-\frac{D}{2}}}{%
\Gamma \left( \frac{D}{2}+1\right) }\int_{0}^{\infty }dy\frac{y^{D}}{%
k_{a}k_{b}e^{2y(b-a)}-1}  \notag \\
&&\times \left[ 1+2D\left( 1-\delta _{1}^{i}\right) (\xi -\xi
_{c})\sum_{j=a,b}k_{j}e^{-2y|\xi -j|}\right] .  \label{Tiiclose}
\end{eqnarray}

In the limit of large proper accelerations for the plates, $a,b\rightarrow 0$%
, for fixed values $a/b$ and $\xi /b$, the world lines of both plates are
close to the Rindler horizon. In this case the single plate and
'interference' parts grow as $b^{-D-1}$. The situation is essentially
different when the world line of the left plane tends to the Rindler
horizon, $a\rightarrow 0$, whereas $b$ and $\xi $ are fixed. By the way
similar to that for the case of the field square, it can be seen that in
this limit the 'interference' part (\ref{intterm1}) vanishes as $\ln
^{-2}(2b/a)$.

In the limit of small proper accelerations, $a,b\rightarrow \infty $ with
fixed values $b-a$, $B_{j}/A_{j}$, and $m$, the main contribution comes from
large values of $\omega $. Using the asymptotic formulas for the Bessel
modified functions, to the leading order one obtains (no summation over $i$)
\begin{eqnarray}
\langle T_{i}^{i}\rangle ^{(j)} &\approx &\frac{4\left( 1-\delta
_{1}^{i}\right) }{(4\pi )^{\frac{D}{2}}\Gamma \left( \frac{D}{2}\right) }%
\int_{m}^{\infty }dy\,(y^{2}-m^{2})^{\frac{D}{2}}  \notag \\
&&\times \frac{e^{-2y|\xi -j|}}{c_{j}(y)}\left( \zeta -\zeta _{c}+\frac{%
\zeta -1/4}{y^{2}-m^{2}}m^{2}\right) ,  \label{EMT1plMink}
\end{eqnarray}%
for the single boundary terms, and%
\begin{eqnarray}
\langle T_{i}^{i}\rangle ^{(ab)} &\approx &-\frac{(4\pi )^{-\frac{D}{2}}}{%
\Gamma \left( \frac{D}{2}+1\right) }\int_{m}^{\infty }dy\frac{(y^{2}-m^{2})^{%
\frac{D}{2}}}{c_{a}(y)c_{b}(y)e^{2y(b-a)}-1}  \notag \\
&&\times \left[ 1-\left( 1-\delta _{1}^{i}\right) \frac{4D(\xi -\xi
_{c})y^{2}-m^{2}}{2(y^{2}-m^{2})}\sum_{j=a,b}\frac{e^{-2y|\xi -j|}}{c_{j}(y)}%
\right] ,  \label{D2Mink0}
\end{eqnarray}%
for the 'interference' term and with the function $c_{j}(y)$ defined by (\ref%
{cjyn}). These expressions are exactly the same as the corresponding
expressions for the geometry of two parallel plates on the Minkowski
background investigated in \cite{Rome02} for a massless scalar and in Ref.
\cite{Mate} for the massive case. In particular, the single boundary terms
vanish for a conformally coupled massless scalar.

In the large mass limit, $ma\gg 1$, by the method similar to that used in
the previous subsection for the field square, it can be seen that the both
single plate and 'interference' parts are exponentially suppressed (no
summation over $i$): $\langle T_{i}^{i}\rangle ^{(j)}\sim m^{D/2+1}\exp
[-2m|\xi -j|]$, $j=a,b$, for single plate parts and $\langle
T_{i}^{i}\rangle ^{(ab)}\sim m^{D/2+1}\exp [2m(a-b)]$ for the 'interference'
part.

\section{'Interaction' forces between the plates}

\label{sec:IntForce}

Now we turn to the 'interaction' forces between the plates due to
the vacuum fluctuations. The vacuum force acting per unit surface of
the plate at $\xi =j$ is determined by the ${}_{1}^{1}$--component
of the vacuum energy-momentum tensor evaluated at this point. The
corresponding effective pressures can be presented as a sum of two
terms:
\begin{equation}
p^{(j)}=p_{1}^{(j)}+p_{\mathrm{(int)}}^{(j)},\quad j=a,b.  \label{FintD}
\end{equation}%
The first term on the right is the pressure for a single plate at
$\xi =j$ when the second plate is absent. This term is divergent due
to the surface divergences in the subtracted vacuum expectation
values and needs additional renormalization. This can be done, for
example, by applying the generalized zeta function technique to the
corresponding mode sum. This procedure is similar to that used in
Ref. \cite{Saha04} for the evaluation of the total Casimir energy in
the cases of Dirichlet and Neumann boundary conditions and in Ref.
\cite{SahSet04b} for the evaluation of the surface energy for a
single Robin plate. This calculation lies on the same line with the
evaluation of the total Casimir energy and surface densities and
will be presented in the forthcoming paper \cite{SahaDav}. Note that
in the formulae for the VEV of the energy-momentum tensor the Robin
coefficients enter in the form of the dimensionless combination
$\beta _j=B_j/(jA_j)$. As a result in the massless case from the
dimensional arguments we expect that the single plate part will have
the form $p_{1}^{(j)}=\alpha (\beta _j)j^{-(D+1)}$. The coefficient
$\alpha (\beta _j)$ in this formula will change if we will change
the renormalization scale and can be fixed by imposing suitable
renormalization conditions which relates it to observables.

The second term on the right of Eq. (\ref{FintD}),
\begin{equation}
p_{\mathrm{(int)}}^{(j)}=-\left[ \langle T_{1}^{1}\rangle ^{(l)}+\langle
T_{1}^{1}\rangle ^{(ab)}\right] _{\xi =j},  \label{pintD}
\end{equation}%
with $j,l=a,b$, $l\neq j$, is the pressure induced by the presence
of the second plate, and can be termed as an 'interaction' force.
This term is finite for all nonzero distances between the plates and
is not affected by the renormalization procedure. Note that the term
'interaction' here should be understood conditionally. The quantity
$p_{\mathrm{(int)}}^{(j)}$ determines the force by which the scalar
vacuum acts on the plate due to the modification of the spectrum for
the zero-point fluctuations by the presence of the second plate. As
the vacuum properties are $\xi $-dependent, there is no a priori
reason for the 'interaction' terms (and also for the total pressures
$p^{(j)}$) to be equal for $j=a$ and $\ j=b$, and the corresponding
forces in general are different. For the plate at $\xi =j$ the
'interaction' term is due to the third summand on the right of Eq.
(\ref{Tik1}). Substituting into this term $\xi =j$ and using the
Wronskian for the modified Bessel functions one has
\begin{equation}
p_{\mathrm{(int)}}^{(j)}=\frac{A_{D}A_{j}^{2}}{2j^{2}}\int_{0}^{\infty
}dk\,k^{D-2}\int_{0}^{\infty }d\omega \left[ \left( \lambda ^{2}j^{2}+\omega
^{2}\right) \beta _{j}^{2}+4\zeta \beta _{j}-1\right] \,\Omega _{j\omega
}(\lambda a,\lambda b),  \label{pint2}
\end{equation}%
with $\beta _j$ defined in the paragraph after formula
(\ref{FintD}). In dependence of the values for the coefficients in
the boundary conditions, the effective pressures (\ref{pint2}) can
be either positive or negative, leading to repulsive or attractive
forces. It can be seen that for Dirichlet boundary condition on
one plate and Neumann boundary condition on the other one has $p_{\mathrm{%
(int)}}^{(j)}>0$ and the 'interaction' forces are repulsive for all
distances between the plates. Note that for Dirichlet or Neumann
boundary conditions on both plates the 'interaction' forces are
always attractive \cite{Avag02}.
By using the relation%
\begin{equation}
\left[ \left( \lambda ^{2}j^{2}+\omega ^{2}\right) \beta _{j}^{2}+\beta
_{j}-1\right] \,A_{j}^{2}\Omega _{j\omega }(\lambda a,\lambda b)=n^{(j)}j%
\frac{\partial }{\partial j}\ln \left\vert 1-\frac{\bar{I}_{\omega
}^{(a)}(\lambda a)\bar{K}_{\omega }^{(b)}(\lambda b)}{\bar{I}_{\omega
}^{(b)}(\lambda b)\bar{K}_{\omega }^{(a)}(\lambda a)}\right\vert ,
\label{pintrel2}
\end{equation}%
with $n^{(j)}$ from (\ref{cjyn}), expressions (\ref{pint2}) for the
'interaction' forces can be written in another equivalent form
\begin{eqnarray}
p_{\mathrm{(int)}}^{(j)} &=&n^{(j)}\frac{A_{D}}{2j}\int_{0}^{\infty
}dk\,k^{D-2}\int_{0}^{\infty }d\omega \left[ 1+\frac{\left( 4\zeta -1\right)
\beta _{j}}{\left( \lambda ^{2}j^{2}+\omega ^{2}\right) \beta _{j}^{2}+\beta
_{j}-1}\right]  \notag \\
&&\times \frac{\partial }{\partial j}\ln \left\vert 1-\frac{\bar{I}_{\omega
}^{(a)}(\lambda a)\bar{K}_{\omega }^{(b)}(\lambda b)}{\bar{I}_{\omega
}^{(b)}(\lambda b)\bar{K}_{\omega }^{(a)}(\lambda a)}\right\vert .
\label{pint3}
\end{eqnarray}%
For Dirichlet and Neumann scalars the second term in the square
brackets is zero. To clarify the dependence of the vacuum
'interaction' forces on the
parameters $a,b$ it is useful to write down the corresponding derivatives:%
\begin{eqnarray}
n^{(j)}\frac{\partial p_{\mathrm{(int)}}^{(j)}}{\partial l} &=&\frac{%
A_{D}A_{a}^{2}A_{b}^{2}}{2abj}\int_{0}^{\infty }dk\,k^{D-2}\int_{0}^{\infty
}d\omega \left[ \left( \lambda ^{2}j^{2}+\omega ^{2}\right) \beta
_{j}^{2}+4\zeta \beta _{j}-1\right]  \notag \\
&&\times \frac{\left( \lambda ^{2}l^{2}+\omega ^{2}\right) \beta
_{l}^{2}+\beta _{l}-1}{Z_{\omega }^{2}(\lambda a,\lambda b)},
\label{pintder}
\end{eqnarray}%
with $j,l=a,b$, $j\neq l$.

Now we consider the limiting cases for the 'interaction' forces
between the plates. For small distances between the plates,
$b/a-1\ll 1$, to the leading order over $1/(b-a)$, the 'interaction'
forces are the same as for the plates
in the Minkowski bulk with the distance $b-a$. The latter are determined by $%
i=1$ component of tensor (\ref{D2Mink0}). In this limit the
'interaction' forces are repulsive in the case of Dirichlet boundary
condition on one plate and non-Dirichlet boundary condition on the
another, and are attractive for all other cases. Note that in the
limit $b\to a$ with fixed values of the boundary coefficients and
the proper acceleration of the left plate, $a^{-1}$, the
renormalized single plate parts $p_1^{(j)}$ remain finite while the
'interaction' part goes to infinity. This means that for
sufficiently small distances between the plates the 'interaction'
term on the right of formula (\ref{FintD}) will dominate.

For large distances between the plates one has $a/b\ll 1$.
Introducing a new integration variable $y=\lambda b$ and using the
asymptotic formulas for the Bessel modified functions for small
values of the argument, we can see that the subintegrand is proportional to $%
(ya/b)^{2\omega }$. It follows from here that the main contribution into the
$\omega $-integral comes from small values of $\omega $. Expanding with
respect to $\omega $, in the leading order we obtain
\begin{subequations}
\label{pintas2gen}
\begin{eqnarray}
p_{\mathrm{(int)}}^{(a)} &\approx &\frac{\pi ^{2}A_{D}\left( 1-4\zeta \beta
_{a}\right) A_{b}^{2}}{24(D-1)a^{2}b^{D-1}\ln ^{3}(2b/a)}\int_{mb}^{\infty
}dy\,\left( y^{2}-m^{2}b^{2}\right) ^{\frac{D-1}{2}}\frac{y^{2}\beta
_{b}^{2}-1}{y\bar{I}_{0}^{(b)2}(y)},  \label{pintas2a} \\
p_{\mathrm{(int)}}^{(b)} &\approx &\frac{\pi ^{2}A_{D}A_{b}^{2}}{%
24b^{D+1}\ln ^{2}(2b/a)}\int_{mb}^{\infty }dy\,y\left(
y^{2}-m^{2}b^{2}\right) ^{\frac{D-3}{2}}\frac{y^{2}\beta _{b}^{2}+4\zeta
\beta _{b}-1}{\bar{I}_{0}^{(b)2}(y)}.  \label{pintas2b}
\end{eqnarray}%
For a massless minimally coupled scalar field these pressures have
the same sign. In Figure \ref{fig2} we have plotted the dependence
of the vacuum 'interaction' forces between the plates as functions
on the ratio $a/b$ for a
massless minimally coupled scalar field in $D=3$ with Robin coefficients $%
\beta _{a}=0$ and $\beta _{b}=1/5$. These forces are repulsive for
small distances and are attractive for large distances. In the
presented example there are parameter choices which give vanishing
'interaction' forces.

\begin{figure}[tbph]
\begin{center}
\epsfig{figure=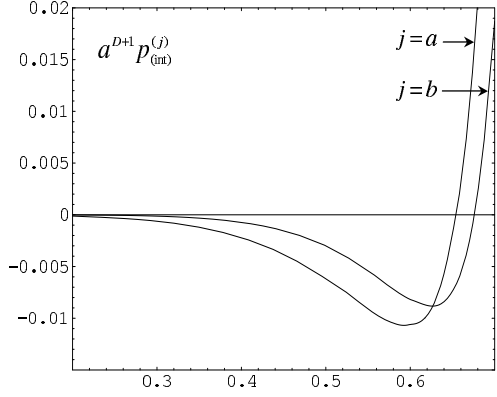,width=7cm,height=5.6cm}
\end{center}
\caption{The vacuum effective pressures $a^{D+1}p_{\mathrm{(int)}}^{(j)}$, $%
j=a,b$, determining the 'interaction' forces between the plates, as
functions on $a/b$ for a massless minimally coupled scalar field in
$D=3$ with Robin coefficients $\protect\beta _{a}=0$ and
$\protect\beta _{b}=1/5$.} \label{fig2}
\end{figure}

For large values of the mass, $ma\gg 1$, the main contribution into the $%
\omega $-integral comes from the values $\omega \sim \sqrt{m(b-a)}$. By
using the uniform asymptotic expansions for the Bessel modified functions,
to the leading order one finds
\end{subequations}
\begin{equation}
p_{\mathrm{(int)}}^{(j)}\approx \frac{B_{j}^{2}m^{\frac{D}{2}+3}e^{2m(a-b)}%
\sqrt{ab}}{(4\pi )^{\frac{D}{2}}c_{l}(m)[A_{j}-n^{(j)}B_{j}m]^{2}j(b-a)^{%
\frac{D}{2}}},  \label{pjlargemass}
\end{equation}%
for $B_{j}\neq 0$, $j,l=a,b$, $j\neq l$. For $B_{j}=0$ the leading term is
the same as that for the $A_{j}=0$ case. The latter is obtained from (\ref%
{pjlargemass}).

\section{Conclusion}
\label{sec:Conc}

The use of general coordinate transformations in quantum field
theory in flat spacetime leads to an infinite number of unitary
inequivalent representations of the commutation relations with
different vacuum states. In particular, the vacuum state for a
uniformly accelerated observer, the Fulling--Rindler vacuum, turns
out to be inequivalent to that for an inertial observer, the
Minkowski vacuum. In the present paper we have considered the
positive frequency Wightman function, the VEVs of the field square
and the energy-momentum tensor for a scalar field in the region
between two infinite parallel plates moving by uniform proper
accelerations, assuming that the field is prepared in the
Fulling-Rindler vacuum state and satisfies the Robin boundary
conditions on the plates. The general case is investigated when
the constants in the Robin boundary conditions are different for
separate plates. The boundaries and boundary conditions are static
in the Rindler coordinates and no Rindler quanta are created. The
only effect of the imposition of boundary conditions on a quantum
field is the vacuum polarization. The Wightman function is
presented in the form of the mode sum involving series over zeros
$\omega =\omega _{n}$ of the
function $Z_{i\omega }(\lambda a,\lambda b)$ defined by relation (\ref%
{Zomega}). For the summation of these series we have applied a summation
formula derived in Appendix \ref{section:App1} by using the generalized
Abel-Plana formula. This allowed to extract from the Whightman function the
part due to a single plate and to present the additional part in terms of
integrals, exponentially convergent in the coincidence limit. The single
plate part is investigated previously in Ref \cite{Saha02}. The contribution
induced by the second boundary is presented in two alternative forms, Eqs. (%
\ref{Wigh3}), (\ref{Wigh31}), obtained from each other by replacements (\ref%
{replacement}). In Section \ref{sec:VEVEMT}, by using the expression
for the Wightman function, we evaluate the VEVs of the field square
and the energy-momentum tensor. The latter is diagonal and the
corresponding components are determined by relation (\ref{Tik1}).
Various limiting cases are studied. In the limit of small distances
between the plates, to the leading order, the VEVs are the same as
those for two parallel plates in the Minkowski vacuum. In the near
horizon limit, $a,b\rightarrow 0$, the proper accelerations of the
plates are large. For fixed values $a/b$ and $\xi /b$, the VEVs grow
as $b^{1-D}$ for the field square and as $b^{-1-D}$ for the
components of the energy-momentum tensor. In the limit when the
world line of the left plate tends to the Rindler horizon,
$a\rightarrow 0$, for a fixed proper accelerations of the right
plate and the observer, the VEVs induced by the left plate vanish as
$\ln ^{-2}(2b/a)$ for both the field square and energy-momentum
tensor. For large values of the mass, the both single plate and
interference parts of the VEVs are exponentially suppressed. The
vacuum forces acting on boundaries are determined by
$_{1}^{1}$-component of the stress and are investigated in Section
\ref{sec:IntForce}. These forces are presented as the sums of two
terms. The first ones correspond to the forces acting on a single
boundary then the second boundary is absent. Due to the surface
divergences in the VEVs of the energy-momentum tensor, these forces
are infinite and need an additional renormalization. The another
terms in the vacuum forces are finite and are induced by the
presence of the second boundary. They correspond to the
'interaction' forces between the plates. These forces per unit
surface are determined by formula (\ref{pintD}). For small distances
between the plates, to the leading order the standard Casimir result
on background of the Minkowski vacuum is rederived. In this limit
the 'interaction' forces are repulsive in the case of Dirichlet
boundary condition on one plate and non-Dirichlet boundary condition
on the another, and are attractive for all other cases. For large
distances, the 'interaction' forces can be either attractive or
repulsive in dependence of the coefficients in the boundary
conditions. In Figure \ref{fig2} we have presented an example when
the vacuum 'interaction' forces are repulsive for small distances
and are attractive for large distances. This provides a possibility
for the stabilization of the interplate distance by vacuum forces.
However, it should be noted that to make reliable predictions
regarding quantum stabilization, the renormalized single plate parts
$p_1^{(j)}$ also should be taken into account. The calculation of
these quantities lies on the same line with the evaluation of the
total Casimir energy and surface densities and will be presented in
the forthcoming paper \cite{SahaDav}.

In the present paper we have investigated the VEV of the bulk
energy-momentum tensor. For scalar fields with general curvature
coupling and Robin boundary conditions, in Ref. \cite{Rome02} it has
been shown that in the discussion of the relation between the mode
sum energy, evaluated as the sum of the zero-point energies for each
normal mode of frequency, and the volume integral of the
renormalized energy density for the Robin parallel plates geometry
it is necessary to include in the energy a surface
term concentrated on the boundary (see also the discussion in Ref. \cite%
{Full03,Milt04}). Similar issues for the spherical and cylindrical boundary
geometries and in braneworld scenarios are discussed in Refs. \cite%
{Saha01,Rome01,Saha04d}. An expression for the surface energy-momentum
tensor for a scalar field with a general curvature coupling parameter in the
general case of bulk and boundary geometries is derived in Ref. \cite%
{Saha04c}. The investigation of the total Casimir energy, the
surface densities, and the energy balance for the geometry under
consideration will be reported in \cite{SahaDav}.

The formulas derived in this paper can be used to generate the vacuum
densities for a conformally coupled massless scalar field in de Sitter
spacetime in presence of two curved branes on which the field obeys the
Robin boundary conditions with coordinate dependent coefficients. The
corresponding procedure is similar to that realized in Ref. \cite{SahSet04}
for the geometry of a single brane and is based on the conformal relation
between the Rindler and de Sitter line elements. The results obtained above
can be also applied to the geometry of two parallel plates near the $D=3$
\textquotedblright Rindler wall.\textquotedblright\ This wall is described
by the static plane-symmetric distribution of the matter with the diagonal
energy-momentum tensor $T_{i}^{k}=\mathrm{diag}(\varepsilon
_{m},-p_{m},-p_{m},-p_{m})$ (see Ref. \cite{Avak01}). Below we will denote
by $x$ the coordinate perpendicular to the wall and will assume that the
plane $x=0$ is at the center of the wall. If the plane $x=x_{s}$ is the
boundary of the wall, when the external ($x>x_{s}$) line element with the
time coordinate $t$ can be transformed into form (\ref{metric}) with
\begin{equation}
\xi (x)=x-x_{s}+\frac{1}{2\pi \sigma _{s}},\quad \tau =2\pi \sigma _{s}\sqrt{%
g_{00}(x_{s})}t.  \label{ksiRw}
\end{equation}%
In this formula the parameter $\sigma _{s}$ is the mass per unit surface of
the wall and is determined by the distribution of the matter:%
\begin{equation}
\sigma _{s}=2\int_{0}^{x_{s}}\left( \varepsilon _{m}+3p_{m}\right) \left[
\frac{g(x)}{g(x_{s})}\right] ^{1/2}dx.  \label{sigmas1}
\end{equation}%
For the \textquotedblright Rindler wall\textquotedblright\ one has $%
g_{22}^{\prime }(x)|_{x=0}<0$ \cite{Avak01} (the external solution for the
case $g_{22}^{\prime }(x)|_{x=0}>0$ is described by the standard Taub
metric). Hence, the Wightman function, the VEVs for the field square and the
energy-momentum tensor in the region between two plates located at $x=x_{1}$
and $x=x_{2}$, $x_{i}>x_{s}$ near the \textquotedblright Rindler
wall\textquotedblright\ are obtained from the results given above
substituting $\xi _{i}=\xi (x_{i})$, $i=1,2$ and $\xi =\xi (x)$. For $\sigma
_{s}>0$, $x\geq x_{s}$ one has $\xi (x)\geq \xi (x_{s})>0$ and the Rindler
metric is regular everywhere in the external region.

\section{Acknowledgements}

The authors are grateful to Armen Yeranyan for useful discussions. This work
was supported by the Armenian National Science and Education Fund (ANSEF)
Grant No. 05-PS-hepth-89-70 and by the Armenian Ministry of Education and
Science Grant No. 0124.

\appendix

\section{Summation formula over zeros of $Z_{iz}(u,v)$}

\label{section:App1}

As we have seen in section \ref{sec:WF}, the Whightman function for a scalar
field in the region between two plates \ uniformly accelerated through the
Fulling-Rindler vacuum is expressed in terms of series over the zeros $%
z=\omega _{n}$ of the function $Z_{iz}(u,v)$, $v>u$, defined by formula (\ref%
{Zomega}). To derive a summation formula for these series, we choose in the
generalized Abel-Plana formula \cite{Sahrev} the functions
\begin{equation}
\begin{split}
f(z)=& \frac{2i}{\pi }\sinh \pi z\,F(z), \\
g(z)=& \frac{\bar{I}_{iz}^{(b)}(v)\bar{I}_{-iz}^{(a)}(u)+\bar{I}%
_{iz}^{(a)}(u)\bar{I}_{-iz}^{(b)}(v)}{Z_{iz}(u,v)}F(z),
\end{split}
\label{DfgtoAP}
\end{equation}%
with a meromorphic function $F(z)$ having poles $z=z_{k}$ ($\neq \omega _{n}$%
), $\mathrm{Im}\,z_{k}\neq 0$, in the right half-plane $\mathrm{Re}\,z>0$.
The zeros $\omega _{n}$ are simple poles of the function $g(z)$. By taking
into account the relation
\begin{equation}
g(z)\pm f(z)=\frac{2\bar{I}_{\mp iz}^{(a)}(u)\bar{I}_{\pm iz}^{(b)}(v)}{%
Z_{iz}(u,v)}F(z),  \label{Dgpmf}
\end{equation}%
for the function $R[f(z),g(z)]$ in the generalized Abel-Plana formula one
obtains
\begin{eqnarray}
R[f(z),g(z)] &=&2\pi i\left[ \sum_{n=1}^{\infty }\frac{\bar{I}_{-i\omega
_{n}}^{(b)}(v)\bar{I}_{i\omega _{n}}^{(a)}(u)}{\frac{\partial }{\partial z}%
Z_{iz}(u,v)|_{z=\omega _{n}}}F(\omega _{n})\right.  \notag \\
&+&\left. \sum_{k}\underset{{z=z_{k}}}{\mathrm{Res}}\frac{F(z)}{Z_{iz}(u,v)}%
\bar{I}_{i\sigma (z_{k})z}^{(b)}(v)\bar{I}_{-i\sigma (z_{k})z}^{(a)}(u)%
\right] ,  \label{DRfg}
\end{eqnarray}%
where the zeros $\omega _{n}$ are arranged in ascending order, and $\sigma
(z_{k})=\mathrm{sgn}(\mathrm{Im\,}z_{k})$. Substituting relations (\ref%
{Dgpmf}) and (\ref{DRfg}) into the generalized Abel-Plana formula, as a
special case the following summation formula is obtained:
\begin{eqnarray}
\sum_{n=1}^{\infty }\frac{\bar{I}_{-i\omega _{n}}^{(b)}(v)\bar{I}_{i\omega
_{n}}^{(a)}(u)}{\frac{\partial }{\partial z}Z_{iz}(u,v)|_{z=\omega _{n}}}%
F(\omega _{n}) &=&\frac{1}{\pi ^{2}}\int_{0}^{\infty }dz\sinh \pi z\,F(z)
\notag \\
&-&\sum_{k}\underset{{z=z_{k}}}{\mathrm{Res}}\frac{F(z)}{Z_{iz}(u,v)}\bar{I}%
_{i\sigma (z_{k})z}^{(b)}(v)\bar{I}_{-i\sigma (z_{k})z}^{(a)}(u)  \notag \\
&-&\int_{0}^{\infty }dz\frac{F(ze^{\frac{\pi i}{2}})+F(ze^{-\frac{\pi i}{2}})%
}{2\pi Z_{z}(u,v)}\bar{I}_{z}^{(a)}(u)\bar{I}_{-z}^{(b)}(v).
\label{Dsumformula}
\end{eqnarray}%
Here the condition for the function $F(z)$ is easily obtained from the
corresponding condition in the generalized Abel--Plana formula by using the
asymptotic formulas for the Bessel modified function and has the form
\begin{equation}
|F(z)|<\epsilon (|z|)e^{-\pi \mathrm{Re\,}z}\left( \frac{v}{u}\right) ^{2|%
\mathrm{Im\,}z|},\quad \mathrm{Re\,}z>0,  \label{condforAPF2pl}
\end{equation}%
for $|z|\rightarrow \infty $, where $|z|\epsilon (|z|)\rightarrow 0$ when $%
|z|\rightarrow \infty $. Formula (\ref{Dsumformula}) can be generalized for
the case when the function $F(z)$ has real poles, under the assumption that
the first integral on the right of this formula converges in the sense of
the principal value. In this case the first integral on the right of Eq. (%
\ref{Dsumformula}) is understood in the sense of the principal value and
residue terms from real poles in the form $\sum_{k}\underset{{z=z_{k}}}{%
\mathrm{Res}}g(z)$, ${\mathrm{Im}\,z}_{k}=0$, have to be added to the
right-hand side of this formula, with $g(z)$ from (\ref{DfgtoAP}). For the
case of series in Eq. (\ref{Wigh1}) the function $F(z)$ is an analytic
function and the residue terms in Eq. (\ref{Dsumformula}) are absent.

\end{document}